# Key Safety Design Overview in AI-driven Autonomous Vehicles

1st Vikas Vyas
*Autonomous Driving Department*
Mercedes-Benz Research and Development North America
Sunnyvale, CA, USA
talkwithvikas@gmail.com

2nd Zheyuan Xu
*Independent Researcher*
University of Washington
Sunnyvale, CA, USA
xuzheyuan961124@gmail.com

*Abstract*—With the increasing presence of autonomous SAE level 3 and level 4, which incorporate artificial intelligence software, along with the complex technical challenges they present, it is essential to maintain a high level of functional safety and robust software design. This paper explores the necessary safety architecture and systematic approach for automotive software and hardware, including fail-soft handling of automotive safety integrity level (ASIL) D (highest level of safety integrity), integration of artificial intelligence (AI), and machine learning (ML) in automotive safety architecture. By addressing the unique challenges presented by increasing AI-based automotive software, we proposed various techniques, such as mitigation strategies and safety failure analysis, to ensure the safety and reliability of automotive software, as well as the role of AI in software reliability throughout the data lifecycle.

## I. INTRODUCTION

The automotive industry continues to change due to self-driving technologies and the involvement of complex software systems. The growth of vehicle automation with minimal user interference necessitates a critical need for safety guidelines and standard implementation. This paper mainly focuses on the safety design features of complex automotive software; however, regarding SAE levels, where the vehicle can perform most of the driving assistance, it still needs driver intervention in many scenarios [1].

The continued increase in autonomous vehicle features adds more complexity and concerns about compliance with safety standards and guidelines. The increasing number of Artificial intelligence algorithms for ADAS functionality like Mapping, perception, and sensor function is creating more issues about performance, predictivity, and reliability, and somehow, it affects the ability to handle unexpected situations [2].

This paper will cover details about current research on key safety design overview in electric and autonomous vehicles, with all the safety strategies and processes. Also, it will address all the important issues and preventive actions to overcome this, such as user interference performance [1], the fail-operational approach [6], [7], and the role of artificial intelligence in safety-critical systems [2], [4]. This article focuses on all the safety standards and guidelines, development, verification, and validation approach to meet the criteria to fulfill the requirement of autonomous driving systems in battery-electric vehicles (BEV) [5], [9], [12].

The paper is divided into the following sections: Section II provides comprehensive details of safety standards and regulations to implement in electric vehicle systems (hardware/software). Section III includes user interference performance and its functions to comply with safety system state transition. Section IV and V reviews the fail-operational and fail-safe methodology to implement in complex automotive software systems. Section VI covers testing, validation, and verification methods for guaranteeing safety. Finally, Section VII summarizes the paper and suggests future research areas.

## II. SAFETY PROPERTIES IN AUTONOMOUS DRIVING

Ensuring safety in autonomous driving systems requires a complex interaction of several safety features. Two key aspects include:

- Functional Safety (FuSa) aims to prevent hazardous occurrences [6], [7], [9] caused by system malfunctions [12], [15], [17]. This includes creating resilient architectures, adding redundancy, and using fault detection and mitigation systems.
- Safety of the Intended Functionality (SOTIF): SOTIF [19] addresses safety problems related to the system's intended functionality [5], [12], even when it performs as designed [16], [17]. This comprises circumstances where the system confronts restrictions due to ambient conditions, sensor limitations, or unexpected edge cases.

On the other hand, ensuring these safety features in L3/L4 autonomous driving systems poses various challenges:

- System Complexity: The complexity of autonomous driving systems, with their complicated interaction of sensors, AI algorithms, and control systems, makes it impossible to detect and mitigate all possible failure scenarios. [12], [14], [16].
- Autonomous vehicles operate in dynamic and unexpected situations, experiencing various road conditions, traffic patterns, and pedestrian behaviors. To ensure safety [3],

[5] in such settings, comprehensive perception, prediction, and decision-making capabilities are required [25], [31], [38].
- User interface: The SAE Levels system requires the user to take control of specific situations. L3/L4 systems need drivers to act in particular scenarios, highlighting the user's attention and immediately taking safety control. In this case, user interference performance has become critical [1].

To conquer this problem, the concept of a safety scenario is building momentum. A safety measure is a systematic argument that can be implemented by review, evidence, and performance to ensure that software is approved for SAE Level for a certain environment [4], [9], [12]. It consists of a systematic analysis of safety hazards, risk analysis, and safety measures to provide a complete assessment of software system performance and measurement.

The criticality of the safety scenario consists of providing evidence of software system safety, complying with international regulations and guidelines, and increasing the user's trust. Also, identifying all unexpected issues and requirement gaps as preventive action during system and software development in electric vehicle development. Continue performing verification and validation to ensure safety improvement and seamless user experience.

### III. From Fail-Safe to Fail-Operational

#### A. Fail operational and fail-safe approach

These are two approaches in automotive software systems, where one is designed to continue functioning with reduced performance after fatal errors in the system. Another side fail-safe strategy helps to lower the risk by taking safety steps to graceful shutdown or safer conditions when any non-fatal or fatal fault occurs in the system [6], [7], [15], such as disabling any ADAS feature like lane keeping, adaptive cruise control, or automated parking control during temporary/parament failure in the system. In Level 3 and Level 4 systems, it is critical that the system continue operating without any interruption from software in case of failure. The fail-operational approach has become a priority in autonomous systems.

#### B. Limitations in Autonomous Driving

Regarding L3 and L4 autonomous driving, this fail-safe approach is insufficient or does not cover all ADAS functions' potential risks and demands. On the Other side, the user may not be fully attentive or ready to take full control of the vehicle because the user always relies on the autonomous system in BEV to handle most driving tasks [1]. BEVs with autonomous software are becoming increasingly normal to interact with other vehicles (V2V) or Vehicles to everything (V2X), like traffic systems and infrastructure. This integration is also becoming more critical for providing an efficient transport system. Meanwhile, many autonomous vehicles and Robotaxis, are forced to stop due to any major failure and transit to a fail-safe state, which can affect the current traffic flow and reduce the efficiency of the complete transportation system.

#### C. Fail-Operational Paradigm

As explained above, this approach concurs with the problem defined in section B by providing continued operation [7], [8] as part of the fault management system. It can reduce the performance (degradation mode) by 50% [10], [11] but ensure the user is not interrupted due to system failure.

#### D. Implications for Software Design

To implement the fail-operational concept, Software architecture must comply with all the safety methods based on safety-level specifications. Here are some of them defined below:
- System Redundancy: Developing redundant subsystems [6], [7] that can dominate the system functioning in the case of a breakdown [10], [11]. This incorporates software redundancies, like algorithms or voting techniques, or hardware redundancies, such as dual-core CPUs or numerous sensors.
- Functional Degradation: The system is designed to transition smoothly to a graceful state, where the functionality is degraded due to significant failure instead of a sudden shutdown. Resource management prioritizes safety-critical functions over non-safety functions. This includes continuing to monitor the entry/exit point via built-in self-test or diagnostic services.
- System Fault Detection and Mitigation: Efficient fault detection and mitigation strategies are crucial for identifying and responding to software or system errors [14], [18], minimizing impact and potential damage, and ensuring system reliability.
- System Adaptation: Implementing an adaptive strategy that allows the system to reconfigure/transition dynamically in case of faults to ensure continued operation. This can involve transiting to a redundant system, changing the control signal, or operating mode.
- System Safety Scenario: Develop complete safety verification and validation cases to demonstrate the system's ability to maintain acceptable safety levels during major failure [4], [9], [12]. This includes detailed failure analysis and impact on safety specification.

Techniques Enabling Fail-Operational Behavior: Many techniques can be used to comply with the fail-operational specifications, some of them are defined below:
- Redundancy of Software function: Developing a different algorithm or monitoring system can help mitigate the impact of Software faults [8], [11].
- Redundancy of hardware function: Utilizing the redundant/backup hardware component, such as dual-core CPU/GPU, sensor, or backup power sources, helps to maintain system availability in case of hardware failure [7], [10], [11].

- Graceful Degradation Strategies: Designing the system to prioritize vital activities while gently degrading non-essential ones in reaction to failures [7], [10].
- Fault-Tolerant Communication: Employing communication protocols that can withstand flaws and retain data integrity, such as CAN with error detection and repair procedures [10].
- Dynamic Reconfiguration: Creating methods to dynamically reconfigure the system in response to errors and change its behavior to retain functioning [10], [11].

Using these methodologies and design ideas, automotive software systems may achieve fail-operational behavior, improving safety, reliability, and availability for autonomous driving systems.

## IV. Safety-Related Availability

In the world of safety-critical automotive systems, availability is a vital attribute [6], [7] that assures the system is operational and ready to execute its intended function when needed [10], [11]. This is especially crucial for autonomous driving systems, which require continuous operation to preserve safety and minimize disturbances to the driving job. High availability requires numerous strategies, such as fault tolerance and redundancy management. Fault tolerance techniques allow the system to function even in the presence of faults or failures, either by disguising the consequences of the problem or by offering alternate means to accomplish the intended functionality [8]. This can be achieved by hardware and software redundancy or a hybrid. For example, redundant sensors may be utilized to cross-check and validate data, while redundant control channels can give alternate ways to activate the vehicle's systems [10], [11].

Redundancy management entails coordinating and controlling the redundant pieces of the system to ensure that they are utilized effectively and efficiently. This involves monitoring the health of redundant components, detecting and isolating errors, and switching to redundant parts as needed [7], [10], [11]. Effective redundancy management is critical for high availability because it guarantees the system can swiftly recover from defects and continue continuous operation. However, there are trade-offs between availability and other safety characteristics, such as safety integrity and performance. Increasing redundancy can enhance availability but also exacerbates system complexity, expense, and possible failure points [11], [12]. Additionally, the added overhead of defect detection and redundancy management might affect system performance, potentially leading to delays or lower responsiveness. As a result, striking the optimal balance between availability and other safety features necessitates careful evaluation of the system's unique requirements and limits. This entails performing a detailed safety analysis to identify possible hazards and risks and then adopting appropriate procedures and tactics to eliminate those risks while maintaining acceptable levels of availability, safety integrity, and performance.

## V. Redundancy and Diversity

Redundancy and diversity are key ideas in safety design. They seek to improve system dependability and resilience by offering various paths to the same objective. Redundancy includes replicating important components or functions, whereas diversity uses alternative technologies or ways to execute the same task.

### A. Principles

- Redundancy: Redundancy: Several instances of a component or function allow the system to continue running even if one fails. This is especially important for safety-critical tasks, as failure might result in dangerous circumstances. As previously explained, redundant braking systems guarantee that the vehicle may still come to a safe stop even if one system malfunctions.
- Diversity: Employing various components or algorithms decreases the possibility of common-cause failures, in which a single event or problem impacts numerous redundant pieces. This can be accomplished using various sensor technologies, software implementations, or alternate control methodologies [13]. For example, integrating camera, radar, and LiDAR sensors offers a more robust and comprehensive representation of the environment, as each sensor has its own strengths and drawbacks.

### B. Examples in Automotive Software

- Redundant Sensors: Numerous sensors of the same or different types are used to achieve overlapping coverage and data cross-validation [20], [22]. This is ubiquitous in perception systems, where redundant cameras, radars, and lidars may increase object recognition and tracking accuracy [31], [33], [34].
- Diverse Algorithms: Implementing numerous algorithms to complete a single task, such as object detection or categorization. This can help offset the impact of algorithmic biases or restrictions since various algorithms may perform better in different contexts. For instance, employing deep learning and rule-based algorithms for object recognition can result in a more robust and adaptive perception system [29], [30].
- Redundant Control Paths: Numerous control routes for activating the vehicle's systems, such as steering and brakes. This ensures the vehicle is still controlled even if one control path fails. To keep vital functions under control, redundant Electronic Control Units (ECUs) or backup communication lines might be employed [7], [17].

### C. Challenges

- Increased Complexity: Implementing and maintaining redundant and diversified components can greatly increase system complexity, posing issues for design, verification, and maintenance. This necessitates careful analysis of trade-offs among redundancy, diversity, and system complexity [14], [16].

- Cost and Resource Overhead: Redundancy and variety frequently have a cost in terms of hardware and software resources. This can affect system or software performance and total hardware cost [11], [12].
- System/Software Integration and Function Coordination: Developing the correct integration and function coordinator function across different systems or software components may be tedious. It requires synchronization, fault management strategies, and resource management [32], [35]
- Developing all safety methods in electric vehicles, such as redundancy, monitoring, and diversity systems, can improve safety levels, reliability, and performance and reduce the risk of challenges.

## VI. Safety Integrity

Safety integrity is critical in automotive systems because it ensures that the system or software implements all the safety specifications/guidelines correctly, even when failure or malfunction occurs [14,15]. To measure safety integrity in automotive systems, we use the Safety integrity level (SIL), which is defined in ISO-26262 [9], [12].

The safety integrity levels (SILS) value is a measure of reliability and availability of safety levels. It is determined based on the frequency and severity of hazards [15], [17]. ISO-26262 defines four SILS levels based on risk, where ASIL A is the lowest level, and ASIL D is the highest level of risk. For each level, we need to define strong safety measures.

### A. Strategy to meet criteria of ASIL requirements:

- Requirement traceability: The requirement traceability is critical for ensuring all the safety requirements are met and comply with the overall product requirements. There are 3 types of requirement traceability – bi-directional traceability, backward traceability, and forward traceability.
- Static Analysis to meet code guidelines: Performing static analysis helps us to avoid any coding issues as well as support meeting safety standards [17].
- Tool Qualification: The tools used to develop and test the system need to be qualified. It is necessary to analyze which tools need qualification and then follow the ISO-26262 guidelines to meet the Qualification criteria.
- Verification and Validation: To ensure system safety, formal validation techniques like unit, integration, software, and system testing are required. Also, Verification method such as review is essential [15], [17].
- Diversity and Redundancy: Implementing redundancy and diversity methods for software and hardware systems, as defined in Section V, support meeting criteria of safety integrity levels by using multiple techniques like PID control, Fuzzy logic, and model predictive control [10]–[12].
- Implementing monitoring, fault detection, and mitigation such as speed monitoring, cross-validation, fallback mode, and driver alert can assist safety integrity even if there is a failure in the system [6]–[8].

### B. AI integration challenge in the System:

To meet safety integrity levels in complex systems, AI-driven system has many challenges:

- Regulatory compliance: Meeting safety regulatory requirements is very challenging for an AI-driven system, especially for a particular software application. Simultaneously, adapting an AI model is also very complex and time-consuming.
- Complexity of AI model: The lack of transparency and complexity can significantly challenge meeting expected specifications and performing verification and validation due to the increased number of model types, such as non-linearity models, black boxes, and adaptive behavior [4], [5].
- Data training and operation: The AI model heavily depends on training and operation, which always impacts data quality and potential results [29], [30], [38].
- Learn and adapt: Continue to learn and adapt is the fundamental characteristic of the AI model for improving their performance and adapting to new challenges, which could result in unintended consequences of safety hazards [38].

Mitigating these problems from AI integration in the Automotive system requires a detailed strategy to integrate the traditional automotive safety approach with innovative techniques from Artificial intelligence. Explainable AI approach [39], which explains the decision-making process for the AI model and its expected impact and potential bias. It uses prediction accuracy and allows validation and monitoring to ensure the safety and reliability of AI-integrated automotive systems.

## VII. Software Components and AI

### A. Perception, fusion, and sensing

All three functions are critical for autonomous driving to achieve Level 3 and Level 4, which enable the autonomous vehicle to develop a 3D model of the surrounding environment that feeds into the vehicle's control system. Sensing gathers all the raw data [21], [23] from sensors such as Lidar, radars, and multiple cameras [24], [26], [31]. Fusion refers to merging the raw data from many sensors and giving output 3D images of the environment (surrounding the vehicle), which helps perception [33], [35]. Perception refers to the processing and interpreting of sensor raw data to detect, identify, and track objects [33], [35]. For this, sensor data quality and processing and interpretation are crucial for complying with safety regulations and specifications in electric vehicles. However, available perception solutions and fusion methods limit their performance, scalability, and reliability, which impacts safety hazards.

When sensor data is not fused, the software does not provide correct inputs from the sensor. The result is that the software is unable to determine the next action. Fusion algorithms have many challenges, such as false or miss detection from sensors, accuracy, and multiple detections from the same sensor [27]. The perception method might have a wrong interpretation of sensor data, which provides incorrect object regeneration, resulting in user errors of Perception, the main factor of accidents [5].

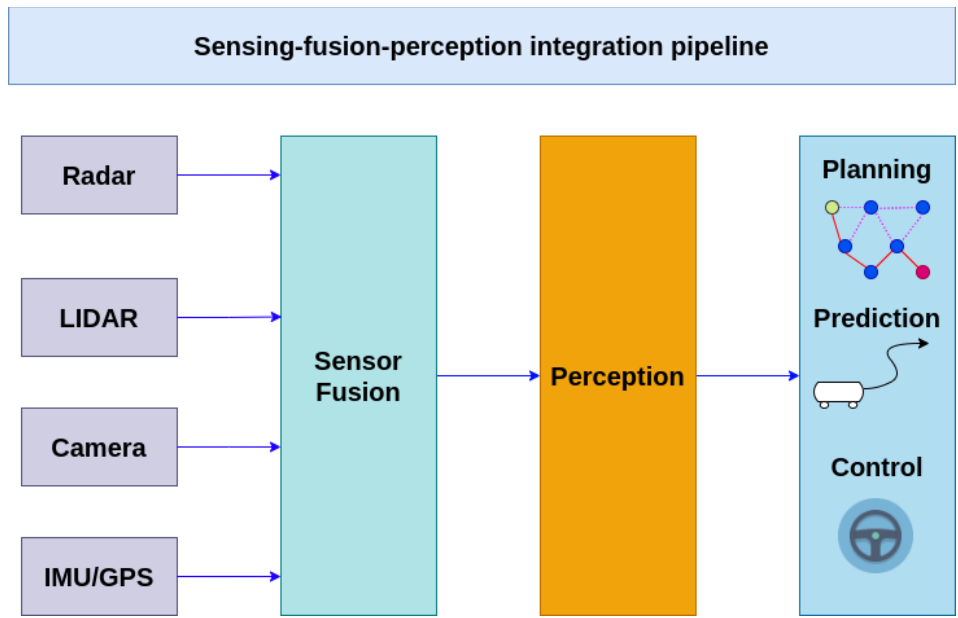

Fig. 1. Sensing, fusion, and perception integration pipeline

Addressing this challenge is crucial for ensuring safety, integrity, and reliability in software in electric vehicles.

Deep learning, one of the AI approaches, uses artificial neuronal networks to learn from data, which improves not only safety and efficiency in automotive, but also fusion, sensing, and perception as mentioned above. Deep learning algorithms recognize the pattern from vast labeled data and provide precise output, which improves object identification [30], [31].

Despite deep learning's numerous advantages, several challenges need to be addressed, such as the massive amount of data needed to train these models and the need for robust and highly implemented deep learning to meet safety specifications and real-world scenarios.

### B. Planning, Prediction, and Control

Planning, prediction, and control are crucial applications for autonomous driving [36], [37] and complying with safety guidelines. Firstly, all possible scenarios of traffic are predicted. Then, the path planning algorithm uses this prediction input to determine a safer path based on current conditions, such as the environment. Finally, the output controls the vehicle from the steering and brake system and informs the driver, if necessary [51], [52]. These applications are critical to implement and integrate seamlessly to ensure automotive safety. However, the current traffic situation can be difficult to predict due to different traffic laws, road geometry, and unpredicted behavior. It uses machine learning algorithms and models [38], [40] to learn from vast driving data and identify patterns and relationships. After training a model, it is crucial to test it for accuracy and effectiveness. Reinforcement learning is a method for training rules for safety [36]. AI-based predictive analysis faces many challenges, such as data quality, quantity, model complexity, and interpretability.

### C. Dataset Lifecycle

Data lifecycle management [53] plays a critical role in ensuring the safety and reliability of AI models used in autonomous electric vehicle applications. This lifecycle includes many states, such as data collection, annotation, storage, cleaning, and balancing.

All AI software applications, especially when used in planning, prediction, control, sensing, perception, and fusion, are critical for safety in autonomous driving in electric vehicles. Figure 2 shows the dataset lifecycle management workflow for Road vehicles as described in ISO-8800. AI integration is a great addition to these applications, but it also increases safety concerns, which need to be properly addressed. Nevertheless, the industry is now seeing a transition towards end-to-end autonomous driving such as [54]–[56], which merges the sensing, perception, and planning through seamless fusion achieved by deep neural networks. While such methods could have more potential compared to the mainstream approach which separates the different modules and treats them as standalone components, the rules and guidelines governing their safety is nearly non-existent.

To summarize, the software components of sensing, perception, fusion, planning, prediction, and control are critical for the safe operation of autonomous vehicles. AI techniques offer significant potential for enhancing these components, but they also introduce new safety challenges that must be carefully addressed. The development of robust and reliable AI-based systems that can handle the complexities and uncertainties of real-world driving scenarios is essential for achieving the full potential of autonomous vehicles while ensuring the safety of all road users.

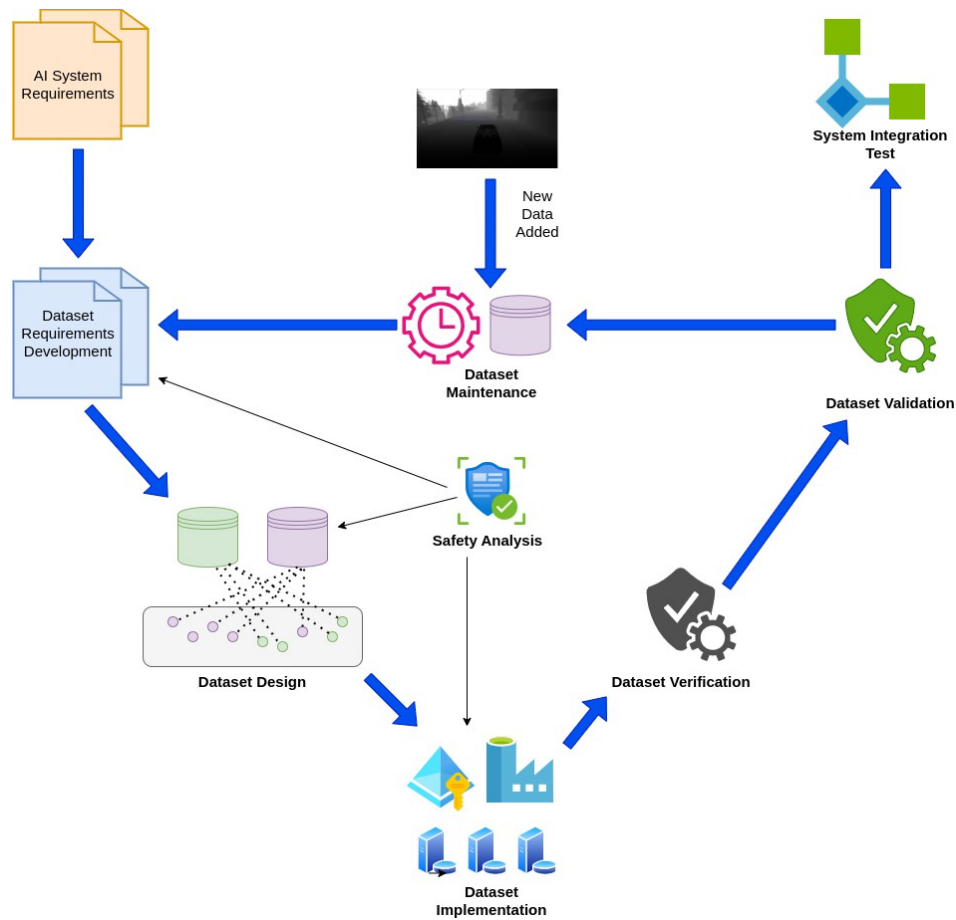

Fig. 2. Dataset lifecycle management workflow as described in ISO-8800 Road vehicles (draft 2022). In real life, dataset management used for training deep learning models deployed in automotive applications involves data collection, annotation, curation, storage, and access control, as well as proper deprecation strategy and consideration for adversarial attacks.

## VIII. CONCLUSION

The automotive sector is rapidly changing due to the adoption of artificial intelligence in driver assistance and autonomous software. This research investigates numerous facets of this technological change, including advances in perception systems, safety mechanisms, sensor fusion approaches, and decision-making algorithms. These developments opened new possibilities in environment perception, object recognition, and motion prediction. Nevertheless, the path to fully autonomous vehicles or level 3 or Level 4 features in electric vehicles is not without challenges, particularly regarding ensuring safety and reliability in complex real-world scenarios. Overcoming these blockers requires a comprehensive approach that includes developing reliable and safer systems, advancing sensing technologies, and conducting thorough testing and validation.

Moreover, the successful adoption of self-driving vehicles relies on establishing standardized frameworks and safety guidelines that encourage industry-wide collaboration and innovation. As research and development in autonomous driving continue to advance, it is crucial to balance technological progress and ethical considerations, ensuring that these innovations benefit society.